# Engineering the spectral profile of photon pairs by using multi-stage nonlinear interferometers

**Mingyi Ma (马明毅)[1], Liang Cui (崔 亮)[1], and Xiaoying Li (李小英)***

*College of Precision Instrument and Opto-Electronics Engineering, Key Laboratory of Opto-Electronics Information Technology, Ministry of Education, Tianjin University, Tianjin 300072, People's Republic of China*
*\*Corresponding author: xiaoyingli@tju.edu.cn*

Using the quantum interference of photon pairs in N-stage nonlinear interferometers (NLI), the contour of joint spectral function can be modified into islands pattern. We perform two series of experiments. One is that all the nonlinear fibers in pulse pumped NLI are identical; the other is that the lengths of N pieces nonlinear fibers are different. We not only demonstrate how the pattern of spectral function changes with the stage number N, but also characterize how the relative intensity of island peaks varies with N. The results, well agree with theoretical predictions in Ref. [1], reveal that the NLI with N pieces nonlinear fibers following binomial distribution can provide a better active filtering function. Our investigation shows that the active filtering effect of multi-stage NLI is a useful tool for efficiently engineering the factorable two-photon state – a desirable resource for quantum information processing.

## 1. Introduction

The mode structure of quantum optical fields is utmost important in quantum information processing (QIP) [2]. The reasons are two folds. First, many protocols in QIP, such as quantum teleportation and quantum computing, are based on quantum interference and it's crucial to have mode match between interfering fields to achieve high visibility [3-5]. Second, the mode profile increases the degrees of freedom for quantum fields, which brings the ability to achieve multidimensional quantum entanglement and increase the capacity of QIP [2,6]. Therefore, many exquisite efforts have been put on engineering the mode structure of quantum states [3,4,7-14].

Engineering the mode structure of quantum state by using nonlinear interferometers (NLI) has recently attracted a lot of attentions [1,15-22]. The NLI, which is analogous to a conventional Mach-Zehnder interferometer but with the two beam splitters being substituted by two parametric amplifiers (PAs), was originally proposed to achieve the Heisenberg limit in precision phase measurement [23] and had been used to demonstrate interesting applications in quantum metrology, spectroscopy and optical imaging etc. [24-27]. When the NLI is used to modify the spectral property of photon pairs, the phase matching of PAs is controlled by the nonlinear medium, while the spectral shaping is obtained through a dispersion control of the interferometer. The separation of the mode control from nonlinear interaction brings the advantages in engineering photon pairs [1] simultaneously possess the features of high purity, high collection efficiency, high brightness, and high flexibility in wavelength and bandwidth selection, and these advantages had been verified by the proof of principle experiments realized by pulse pumped two- and three-stage NLIs [19-21].

The theoretically analysis shows that finer mode control can be realized if the stage number of NLI, N (the number of PAs or nonlinear media), is greater than 2 [1]. For the joint spectral function (JSF) of signal and idler photon pairs with islands pattern tailored by the active filtering effect of NLI, the separation between two adjacent islands increases with N. When the gain of each PA is the same, the interference factor of the JSF, $\frac{\sin N\theta}{\sin \theta}$ (with $\theta$ denoting the phase shift induced by dispersion control in NLI), is similar to that of a multi-slit interferometer in classical optics. In this case, there are mini-maxima existed in between the main maxima, leading to non-ideal isolation between two adjacent islands. For the frequency uncorrelated photon pairs extracted from one island, the non-ideal isolation results in reduced collection efficiency [19-21]. When the gains of PAs are changed by properly arranging the lengths of nonlinear media, the influence of the mini-maxima on the isolation between adjacent islands can then be eliminated. However, most of the NLI experiments presented so far [19-22,24-27] consists of nonlinear media with identical length. Modifying the JSF by using multi-stage NLI with uneven length nonlinear media has not been experimentally demonstrated yet.

In this paper, we experimentally investigate the JSF of photon pairs generated from four wave mixing (FWM) in pulse pumped multi-stage NLIs, which are formed by a sequential array of nonlinear fibers, with a gap in between made of a linear dispersive medium of standard single mode fiber. To illustrate that the multi-stage NLIs consisting of nonlinear fibers with uneven length have the advantages in providing a better active filtering function for reshaping the JSF, we perform two series of experiments and compare their results. One is that all the nonlinear fibers in NLI with $N = 2,3,4$ are identical; the other is that the lengths of $N(N>2)$ pieces nonlinear fibers follow binomial distribution.

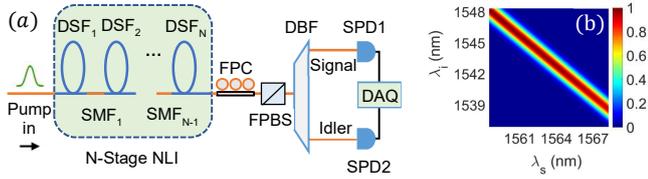

Fig. 1. (a) The experimental setup of generating photon pairs from an $N$-stage nonlinear interferometer (NLI) consisting of N pieces nonlinear media of dispersion shifted fibers (DSFs) and $N-1$ pieces phase shifters of standard single mode fibers (SMFs). DBF, dual-band filter; FPC, fiber polarization controller; FPBS, fiber polarization beam splitter; SPD, single photon detector; DAQ, data acquisition system. (b) The contour of joint spectral intensity $|F(\omega_s, \omega_i)|^2$ for photon pairs generated from single piece DSF.

## 2. Experiments and results

Our experimental setup is shown in Fig. 1. The NLI is formed by $N$ pieces of dispersion shifted fibers (DSFs) with $N-1$ pieces of standard single mode fiber (SMFs) in between. The DSF functions as nonlinear medium of FWM; the SMF functions as linear dispersion medium and is used to introduce phase shift in NLI. Since the spatial mode of optical fields involved in FWM is well confined by the waveguide structure of optical fibers, we can focus on studying the temporal or spectral mode structure of photon pairs. For each DSF, the zero dispersion wavelength is $\lambda_0 = 1552.5$ nm, the dispersion slope at $\lambda_0$ is $D_{slope} = 0.075$ ps/(km·nm$^2$), and the nonlinear coefficient $\gamma$ is $2$ (W·km)$^{-1}$. When the NLI is pumped with a train of laser pulses with central wavelength of $\lambda_{p0} = 1553.3$ nm, the phase matching condition of FWM, $\Delta k = \frac{\lambda_{p0}^2}{8\pi c} D_{slope}(\lambda_{p0} - \lambda_0)(\omega_s - \omega_i)^2 - 2\gamma P_p \approx 0$ is satisfied [21], where $c$ is the speed of light in vacuum, $P_p$ is the peak pump power. In the spontaneous FWM process, two pump photons at frequency $\omega_p$ scatter through the $\chi^{(3)}$ Kerr nonlinearity of DSF to create a pair of quantum correlated signal and idler photons at frequencies of $\omega_s$ and $\omega_i$ respectively. The dispersion coefficient of standard SMF in 1550 nm band is about $D_{SMF} = 17$ ps/(nm·km). So the phase matching condition of FWM is not satisfied in SMF. The spectral shaping of photon pairs is obtained through the control of linear dispersion media of SMF in NLI.

The pump of NLI is obtained by passing the output of a femto-second fiber laser through a bandpass filter (see Ref. [19] for details). The full-width at half maximum (FWHM) and repetition rate of transform limited pump pulses are 1.4 nm, and 36.8 MHz, respectively. In the NLI, the transmission losses of DSFs and SMFs, which are about 0.3 dB/km 0.2 dB/km, respectively, are negligible and the loss induced by each splicing point between DSF and SMF is less than 4%. In the experiments presented hereinafter, the length of each SMF, $L_{DM}$, is fixed at 10 m, while the number and length of DSFs can be varied.

Since the internal loss of our NLI is negligibly small, when the pump power is low and the length of each DSF is identical, the JSF of photon pairs at the output of NLI can be approximately expressed as [1]:

$$F_{NLI}^S(\omega_s, \omega_i) = F(\omega_s, \omega_i) \times H(\theta)$$
$$= g \exp\left[-\frac{(\omega_s + \omega_i - 2\omega_{p0})^2}{4\sigma_p^2}\right] \mathrm{sinc}\left(\frac{\Delta k}{2}\right) \times H(\theta) \quad (1)$$

where $F(\omega_s, \omega_i) = g \exp\left[-\frac{(\omega_s + \omega_i - 2\omega_{p0})^2}{4\sigma_p^2}\right] \mathrm{sinc}(\frac{\Delta kL}{2})$ with gain coefficient $g \propto \gamma P_p L \ll 1$ is the JSF of photon pairs generated by single piece DSF, the superscript "S" denotes that the length of each DSF in NLI is the same, $\omega_{p0} = 2\pi c/\lambda_{p0}$ and $\sigma_p$ are the central frequency and bandwidth of pump. In Eq (1),

$$H(\theta) = e^{i(N-1)\theta} \frac{\sin N\theta}{\sin \theta}, \quad (2)$$

is the interference factor of NLI, where $\theta = (\Delta kL + \Delta\phi_{DM})/2$ with $\Delta\phi_{DM}$ representing the phase shift induced by SMF is the overall phase difference between pump and photon pairs of adjacent DSFs. In our experiment, the approximations $\Delta kL \to 0$ and $\mathrm{sinc}\left(\frac{\Delta kL}{2}\right) \to 1$ hold due to the satisfaction of phase matching condition in DSF. For the SMF with $L_{DM} = 10$ m, the amount of phase shift $\Delta\phi_{DM}$ is much greater than that of the term $\Delta kL$, and we have the approximation [19,21]:

$$\theta \approx \frac{\Delta\phi_{DM}}{2} = \frac{\lambda_{p0}^2 D_{SMF} L_{DM}(\omega_s - \omega_i)^2}{16\pi c}. \quad (3)$$

When the linear dispersion media of SMFs still have the same length but the lengths of DSFs are arranged by using binomial distribution [1]:

$$L_n = L_1 C_{N-1}^{n-1} = L_1 \frac{(N-1)!}{(n-1)!(N-n)!}, \quad (4)$$

the JSF of photon pairs can then be approximately expressed as [1]:

$$F_{NLI}^{(UN)}(\omega_s, \omega_i) \approx \exp\left[-\frac{(\omega_s + \omega_i - 2\omega_{p0})^2}{4\sigma_p^2}\right] \times K(\theta), \quad (5)$$

with the interference factor given by

$$K(\theta) = \sum_{n=1}^N L_n e^{2i(n-1)\theta} = L_1\left(1 + e^{2i\theta}\right)^{N-1}$$
$$= L_1 2 \cdot 4^{(N-1)} \cos^{2(N-1)}\theta, \quad (6)$$

where $L_n$ is the length of DSF$_n$ (n = 1,2,…N) in NLI (see Fig. 1(a)), the superscript "UN" refers to the NLI with uneven length DSFs, $\theta$ is the same as in Eq. (3). Therefore, if the different sections of nonlinear media in NLI follow a binomial pattern, e.g., $L_1:L_2:L_3 = 1:2:1$ for $N = 3$ and

$L_1: L_2: L_3: L_4 = 1: 3: 3: 1$ for $N = 4$, there is no mini-maxima in $|K(\theta)|^2$. On the contrary, the H-function in Eq. (2) normally gives $(N − 1)$-th harmonic of $\cos 2\theta$, i.e., $\cos 2(N − 1)\theta$.

To experimentally characterize the spectral profile of photon pairs, we need to extract the signal and idler photon pairs at the output of NLI. A fiber polarization controller (FPC) placed in front of the fiber polarization beam splitter (FPBS) is used to select the signal and idler photon pairs co-polarized with the pump and to reject the Raman scattering (RS) cross-polarized with the pump [28]. Because the conversion efficiency of FWM in the NLI is relatively low, about 0.01 photon pair is produced in a 100-m-long DSF when the number of pump photons contained in a pulse of 4 ps duration is about $10^7$. Thus, to reliably detect the correlated photon pairs by single photon detectors (SPDs), a pump to photon-pair rejection ratio in excess of 110 dB is required. We achieve this by passing the output of NLI through a dual-band filter (DBF), which is realized by cascading a notch filter with a programmable optical filter (POF, model: Finisar Wave shaper 4000S). For each pass band of DBF, the spectrum is rectangularly shaped, both the central wavelength and bandwidth are tunable.

Two superconducting nanowire single photon detectors SPD1 and SPD2 followed by a data acquisition system (DAQ) are utilized to count the signal and idler photons. The detection efficiencies in both signal and idler bands are about 10% when the efficiencies of SPDs (~ 80%) and transmission efficiency of DBF are included.

We first measure the JSF of photon pairs generated from NLI when the DSFs are the same and the stage number is $N = 2, 3, 4$, respectively. In the experiment, the length of each DSF is 100 m, the average pump power is fixed at 60 μW, the bandwidth for both the signal and idler pass bands of DBF is set to ~ 0.16 nm (0.02 THz in frequency). During measurement, the central wavelength of DBF in signal (idler) channel is scanned from 1558.5 nm (1548.4 nm) to 1568.3 nm (1537.9 nm) with a step of ~ 0.16 nm (0.02 THz in frequency). Under each wavelength setting of DBF, we not only record the single counts of individual SPD1 and SPD2, respectively, but also measure the two-fold coincidence counts of two SPDs respectively for the signal and idler photons originated from the same pump pulse and adjacent pulses, $C_c$ and $C_{acc}$. We then deduce the true coincidence counts of photon pairs $C_T$ by subtracting the measured $C_{acc}$ from $C_c$. Figure 2(c) plots the contour maps of true coincidences in the wavelength coordinates of $\lambda_s$ and $\lambda_i$, which reflects the joint spectral intensity (JSI) $|F_{NLI}^S(\omega_s, \omega_i)|^2$ of photon pairs [29,30]. Clearly, the contour maps of true coincidence for the cases of $N = 2, 3, 4$ in Fig. 2(c) exhibit islands pattern, which are obviously different from the JSI of photon pairs produced by a signal piece DSF (see Fig. 1(b)). For each plot in Fig. 2(c), the distance between the maxima of adjacent primary islands decreases with the increase of detuning ($\lambda_s − \lambda_i$) because the phase shift $\theta$ in Eq. (3) quadratically depends on the detuning of photon pairs. For the convenience of recognition, we label the primary islands with $m = 1, 2, 3, 4$, and the order number $m$ increases with the detuning ($\lambda_s − \lambda_i$). From Fig. 2(c), one sees that the variation trend for patterns of JSF has the following distinct features. First, the central wavelengths of the primary islands do not vary with $N$. In each plot of Fig. 2(c), the central wavelengths of the islands labelled $m = 1, 2, 3, 4$ in the signal (idler) band are centering at about 1560.4 nm (1546.3 nm), 1563.3 nm (1543.5 nm), 1565.5 nm (1541.3 nm), 1567.4 nm (1539.5 nm), respectively. Second, for the island with a fixed order number $m$, the width of the island decreases with the increase of $N$. Accordingly, the separation between adjacent islands increases with $N$. Third, the peak intensity for the measured true coincidences of the NLI increases with $N$. The counting rates corresponding to the highest peaks of the primary islands are 87±5, 205±10 and 323±18 counts/s for $N = 2, 3, 4$, respectively. The results show that the peak of island increases with the scale of $N^2$, as predicted by Eq. (2). Fourth, there exists $N − 2$ secondary islands between two adjacent primary islands. The counting rates for all the peaks of the secondary islands are around 20-30 counts/s. So the ratio between the intensities of the primary and secondary islands increases with the increase of $N$. As a result, the influence of secondary islands upon the isolation between two primary islands will decrease for $N$ with a larger number [15].

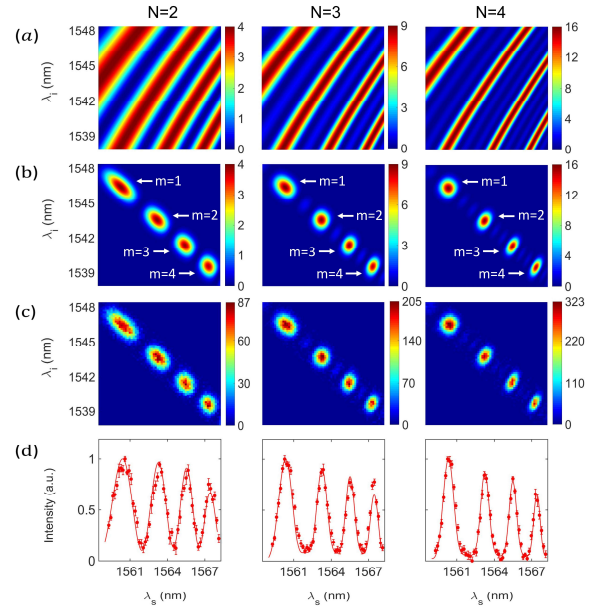

Fig. 2. The results for $N$-stage NLI with $N = 2, 3, 4$ when the lengths of each DSF and SMF in NLI are 100 m and 10 m, respectively. (a, b) The contours of interference factor $|H(\theta)|^2$ and joint spectral intensity $|F_{NLI}^S(\omega_s, \omega_i)|^2$ calculated by substituting experimental parameters into Eqs. (2) and (1), respectively. (c) Contour maps of measured true coincidence of photon pairs, which reflect the joint spectral intensity of photon pairs. (d) Measured (dots) and calculated (solid curves) marginal intensity distributions of signal photons.

To compare the measured JSIs with theoretical predictions, we calculate the contours of JSI for the case of $N = 2, 3, 4$ by substituting the experimental parameters of our NLI into Eqs. (1)-(3), as shown in Fig. 2(b). To better illustrate how the quantum interference of photon pairs in NLI influences the JSF, the contours of interference factor $|H(\theta)|^2$ for $N = 2, 3, 4$ are calculated as well, as shown in Fig. 2(a). One sees that the symmetry lines of $|H(\theta)|^2$ in Fig. 2(a) are perpendicular to that of $|F(\omega_s, \omega_i)|^2$ in Fig. 1(b). As a result, the JSIs of NLI in Fig. 2(b) exhibit islands pattern and the main maxima occur at $\theta = m\pi$ ($m = 1, 2, \dots$). Depending on the stripe width of $|H(\theta)|^2$, $\sigma_{int}^2 = 4c/[(N-1)m\lambda_{p0}^2 D_{SMF} L_{DM}]$, the pattern of primary island changes with variations of the stage number $N$, order number of island $m$, and pump bandwidth [1,21]. With the increase of $N$ and $m$, the width of the primary island decreases. The comparison of Figs. 2(c) and 2(b) shows that the experimental results well agree with the theoretical predictions.

It's worth noting that the multi-stage NLI with $N(N > 2)$ pieces identical nonlinear media had already been experimentally realized in fiber-based NLI and $\chi^{(2)}$ crystal-based NLI [20,21,27], respectively. In particular, in Ref. [27], the JSIs of NLI were measured when the stage number was $N = 2,3,4,5$, respectively, while other experimental parameters were fixed. Although the results qualitatively agree with theory, there are obvious deviation from theoretical predictions due to the problem of imperfect alignment in bulk-crystal based system [27]. Here, we characterize the islands pattern of JSI from top to bottom by using the advantages of the fiber-based NLI: freedom from misalignment and low internal loss for all the optical fields involved in FWM. In addition to illustrating how the center and width of islands change with the stage number $N$ and order number $m$, we reveal how the relative intensity of island peaks varies with $N$. The latter has not been demonstrated before.

In order to clearly visualize that the NLI functions as an active filter of photon pairs, we deduce the marginal intensity distribution in signal and idler fields from the counting rates of SPD1 and SPD2, respectively. Because the data of the two individual fields is similar, for the sake of brevity, we only show the results in signal field. Note that for our fiber-based NLI, there is Raman scattering (RS) accompanying the FWM [28]. The existence of RS does not affect the measured JSI in Fig. 2, because there is no quantum correlation for the RS photons in signal and idler bands. However, when the active filtering effect of NLI is characterized by observing the interference in individual signal (idler) band, which is equivalent to projecting the contour of JSI along signal (idler) axis, the contribution of RS should be subtracted.

When the pass bands of DBF are scanned, at each wavelength setting, we measure the counting rates $R_{s(i)}$ in individual signal (idler) channel at different levels of average pump power $P_a$ and then fit the measured data with a second-order polynomial $R_{s(i)} = s_1 P_a + s_2 P_a^2$, where $s_1$ and $s_2$ are the fitting parameters. The linear and quadratic terms, $s_1 P_a$ and $s_2 P_a^2$, are respectively proportional to the intensities of RS and FWM. Figure 3 shows two typical sets of raw data, which are obtained by setting the wavelength of DBF in signal band at 1560.4 nm and 1561.9 nm, respectively. The two wavelengths respectively correspond to the peak and valley of the islands with $m = 1$ and $N = 4$ (see Fig. 2(c)). The fitting results of the data show that the linear parts in both Figs. 3(a) and 3(b) are quite high. This is because our NLI is simply placed at room temperature to avoid unnecessary complexity. Considering the RS can be significantly suppressed by cooling NLI [19,21,31], here we focus on studying the photon pairs via FWM. Comparing with the quadratic part (dotted curve) in Fig. 3($a$), we find that the value of quadratic term in Fig. 3(b) is quite small due to the destructive interference effect occurred at the wavelength of ~1561.9 nm in NLI with $N = 4$. When pump power is 60 μW, the ratio between the quadratic terms in Figs. 3($a$) and 3(b) is about 16.3, indicating the visibility of interference fringe individual signal field originated from FWM is around 88%, which is slightly lower than the theoretically calculated result of ~92%. We think the deviation from theoretical prediction is caused by the internal loss in NLI (~15% for $N = 4$) due to the imperfect splicing [32].

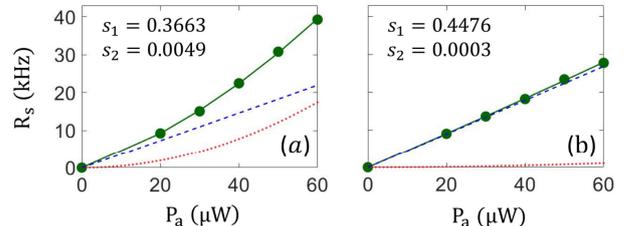

Fig. 3. Measured counting rate (dots) $R_s$ as a function of the average pump power $P_a$ for the photons in individual signal band centering at ($a$) 1560.4 nm and (b) 1561.9 nm, respectively. The solid curve $R_s = s_1 P_a + s_2 P_a^2$ is the fitting function of raw data (dots), dashed and dotted lines are the linear term $s_1 P_a$ and quadratic term $s_2 P_a^2$, respectively.

When the interference fringe in individual signal (idler) field is used to characterize the active filtering effect of NLI, we only consider the contribution of photons originated from FWM, which are deduced from quadratic part of the second-order polynomial fitting function. Fig. 2(d) shows the normalized rate of single counts (via FWM) as a function of the wavelength in signal channel. Each plot in Fig. 2(d) exhibits interference pattern, illustrating the NLI functions as an active filter of photon pairs. The data (dots) in Fig. 2(d) is obtained by subtracting the RS from the directly measured counts of SPD1, as illustrated in Fig. 3. We find the measured intensity distributions agree with the theoretically calculated results (solid curves). For the interference fringes correspond to a fixed peak wavelength, which are originated from the islands with same order number $m$, the fringe width decreases with the increase of $N$. The visibility of the interference fringe, defined by $V_m = (I_{max}^m - I_{min}^m)/(I_{max}^m + I_{min}^m)$ with $I_{max}^m$ and $I_{min}^m$ respectively denoting the normalized intensity at the peak and trough, increases with $N$. One sees that $V_m$ is the lowest for the case of $N = 2$ due to the existence of overlap of between two adjacent islands. However, even if $N$ is increased to a larger number, the minimum of the normalized intensity (single count rates) $I_{min}^m$ is still away from 0 due to the secondary islands existed in between adjacent primary islands. Note that in the process of selecting the photon pairs with spectral profile determined by one specific island, the bandwidth of filters used to efficiently select out photon pairs should be properly set. By doing so, one island can be picked out as a whole and the isolation to adjacent islands is as much as possible [21]. Hence, the collection efficiency of photon pairs, characterized by the probability of detecting the photon at signal (idler) band for a photon detected in the idler (signal) band, is closely related to interference fringe presented in marginal intensity distribution. To improve the collection efficiency, the visibility of the corresponding fringe in Fig. 2(d) should be as high as possible. The non-ideal visibility (less than 1) will prevent the collection efficiency of photon pairs from reaching the ideal value of 1, in particular, when the NLI is designed to engineer the factorable state [1,19,21].

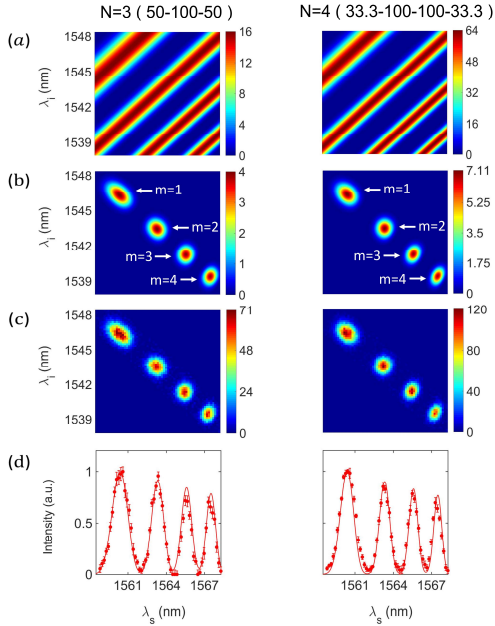

Fig. 4. The results for uneven $N$-stage NLI, in which the lengths of DSFs, labelled in the top for different stage number $N$, follow binomial distribution. (a,b) The calculated contours of interference factor $|K(\theta)|^2/(L_1^2)$ and joint spectral intensity $|F_{NL}^{(UN)}(\omega_s, \omega_i)|^2$ (see Eqs. (5)-(6)), respectively. (c) Contour maps of true coincidence of photon pairs, reflecting the joint spectral intensity of photon pairs. (d) Measured (dots) and calculated (solid curves) marginal intensity distributions of signal photons.

To illustrate how to get rid of the mini-maxima in between the main maxima in the JSI in Fig. 2, we then perform the second experiment to verify the spectral profile of uneven $N$-stage NLI with $N = 3, 4$, respectively. In the experiment, the parameters are the same as that in Fig. 2, except for the lengths of DSFs, which follow binomial distribution (see Eq. (4)). When $N = 3$, the lengths of three DSFs, $L_1$, $L_2$, and $L_3$ are 50, 100 and 50 m, respectively; when $N = 4$, the lengths of four DSFs, $L_1$, $L_2$, $L_3$ and $L_4$, are 33.3, 100, 100, and 33.3 m, respectively. Figure 4(c) shows the contour maps of true coincidences in the wavelength coordinates of signal and idler, $\lambda_s$ and $\lambda_i$. The counting rates corresponding to the highest peaks of the islands are 71±3 and 120±5 counts/s for $N = 3, 4$, respectively. The centers of islands and the variation tendencies of the islands are the same as those of $N = 3$ and $N = 4$ cases in Fig. 2. The separation between two adjacent islands still increases with $N$, but the secondary islands do not exist anymore.

We also calculate the contours of interference factor $|K(\theta)|^2/(L_1^2)$ and JSI $|F_{NL}^{(UN)}(\omega_s, \omega_i)|^2$ of uneven $N$-stage NLI by substituting the experimental parameters into Eqs. (5)-(6), as shown in Fig. 4(a) and 4(b). It's clear that the variation tendency of $|K(\theta)|^2/L_1^2$ is similar to that in Fig. 2(a), except that the mini-maxima does not exist anymore. The stripe width of the main maxima of $|K(\theta)|^2/(L_1^2)$ still narrows as $N$ increases, leading to enlarged islands separation in JSF. Comparing Fig. 4(c) with Fig. 4(b), we find the measured islands pattern of JSI well agrees with theory prediction.

Additionally, we characterize the active filter function of the uneven $N$-stage NLI by deducing the marginal intensity distribution in individual signal and idler fields. The procedure of processing data is similar to that in Figs. 2(d) and 3. Fig. 4(d) shows the data (dots) in signal channel. We find the data agrees with the theoretically calculated results (solid curves). Comparing with Fig. 2(d), One sees that for the NLI with same stage number of $N$, the visibility of interference fringes in Fig. 4(d) is slightly higher.

It's well known that the spectrally factorable photon pairs with high collection efficiency are the desirable resources of quantum information processing. Recent investigation shows that the quantum interference in NLI provide an alternative approach for achieving this kind of photon pairs [1,21,22]. So, we pay more attention on the round shaped islands, from which the spectrally factorable state can be obtained without sacrificing collection efficiency. For the round islands in Fig. 2, obtained under the condition of (i) $N = 2, m = 4$, (ii) $N = 3, m = 2$, and (iii) $N = 4, m = 1$, the visibility in Fig. 2(d) obviously deviate from the ideal results of 1 due to the existence of overlap ($N = 2$) or the mini-maxima ($N \geq 3$). While for both the cases of (ii) $N = 3$, $m = 2$ and (iii) $N = 4$, $m = 1$ in Fig. 4, the fringe visibilities of the round island in Fig. 4(d) are approaching to the ideal value of 1 because of the increased separation and the elimination of mini-maxima. This analysis and comparison indicate that if the NLI is used to engineering factorable state, higher collection efficiency of photon pairs can be obtained by the properly designed uneven $N$-stage NLI with $N \geq 3$.

### 3. Summary and Discussion

In conclusion, we have experimentally investigated the spectral profile of photon pairs generated from pulse pumped multi-stage nonlinear interferometers, in which DSFs and standard SMFs respectively function as nonlinear fibers and linear dispersion media. Since the fiber-based NLI has the advantages of freedom from misalignment and low internal loss, we are able to comprehensively characterize the joint spectral function of photon pairs. The experimental results agree with theoretical predictions. Our investigation shows that although the photon pairs produced by single piece DSF via four wave mixing are frequency anti-correlated, their mode profile can be flexibility modified by active filtering effect originated from the quantum interference in $N$-stage NLI, and the NLI with stage number $N > 2$ has more flexibility in modifying the mode structure of photon pairs.

Moreover, the experimental results reveal that the uneven multi-stage NLI can provide a better active filtering function: the separation between adjacent islands in the contour of joint spectral function increases with $N$, and there is no mini-maxima between two primary maxima. To the best of our knowledge, this is the first experimental demonstration of using uneven multi-stage NLI to modify JSF. In principle, the perfect active filtering effect is achievable by increasing $N$ to a large number, even if each nonlinear medium in NLI is identical [15]. However, in practice, the increased stage number usually accompanies increased internal loss of NLI, which will not influence the islands pattern of JSF for fiber-based NLI but introduce uncorrelated background noise photons in individual signal and idler band [32]. Therefore, the feature of the uneven multi-stage NLI is useful for efficiently engineering the factorable two-photon state, in particular, when the stage number of NLI is not large.

This work was supported by the National Natural Science Foundation of China (11527808, 11874279), and Science and Technology Program of Tianjin (18ZXZNGX00210).

[1]These authors contributed equally to this work.

### References
1. L. Cui, J. Su, J. Li, Y. Liu, X. Li, and Z. Y. Ou, Phys. Rev. A, 102, 033718 (2020)
2. C. Fabre and N. Treps, Rev. Mod. Phys, 92, 035005 (2020).
3. J. Pan, D. Bouwmeester, W. Harald, and Z. Anton, Phys. Rev. Lett. 80, 3891 (1998).
4. D. Bouwmeester, J. Pan, M. Klaus, E. Manfred, W. Harald, and Z. Anton, Nature 390, 575 (1997).
5. E. Knill, R. Laflamme, and G. J. Milburn, Nature 409, 46 (2001).
6. B. Brecht, Dileep V. Reddy, C. Silberhorn, and M. G. Raymer, Phys. Rev. X 5, 041017 (2015)
7. Z. Y. Ou, Quantum Semiclass. Opt. 9, 599 (1997).
8. W. P. Grice and I. A. Walmsley, Phys. Rev. A 56, 1627 (1997).
9. P. J. Mosley, J. S. Lundeen, B. J. Smith, P. Wasylczyk, A. B. U' Ren, C. Silberhorn, and I. A. Walmsley, Phys. Rev. Lett. 100, 133601 (2008).
10. K. Garay-Palmett, H. J. McGuinness, O. Cohen, J. S. Lundeen, R. Rangel-Rojo, A. B. U'Ren, M. G. Raymer, C. J. McKinstrie, S. Radic, and I. A. Walmsley, Opt. Express 15, 14870 (2007).
11. O. Cohen, J. S. Lundeen, B. J. Smith, G. Puentes, P. J. Mosley, and I. A. Walmsley, Phys. Rev. Lett. 102, 123603 (2009).
12. L. Cui, X. Li, and N. Zhao, New. J. Phys. 14, 123001 (2012).
13. P. G. Evans, R. S. Bennink, W. P. Grice, T. S. Humble, and J. Schaake, Phys. Rev. Lett. 105, 253601 (2010).
14. A. M. Brańczyk, A. Fedrizzi, T. M. Stace, T. C. Ralph, and A. G. White, Opt. Express 19, 55 (2011).
15. A. B. U' Ren, C. Silberhorn, K. Banaszek, I. A. Walmsley, R. Erdmann, W. P. Grice, and M. G. Raymer, Laser Phys. 15, 146 (2005).
16. G. Frascella, E. E. Mikhailov, N. Takanashi, R. V. Zakharov, O. V. Tikhonova, and M. V. Chekhova, Optica 6, 1233 (2019).
17. G. Frascella, R. V. Zakharov, O. V. Tikhonova, and M. V. Chekhova, Laser Phys. 29, 124013 (2019).
18. S. Lemieux, M. Manceau, P. R. Sharapova, O. V. Tikhonova, R. W. Boyd, G. Leuchs, and M. V. Chekhova, Phys. Rev. Lett. 117, 183601 (2016).
19. J. Su, L. Cui, J. Li, Y. Liu, X. Li, and Z. Y. Ou, Opt. Express 27, 20479 (2019).
20. J. Li, J. Su, L. Cui, X. Li and Z.Y. Ou, Conference on Lasers and Electro-Optics, 2019, FTh3D.5.
21. J. Li, J. Su, L. Cui, T. Xie, Z. Y. Ou, and X. Li, Appl. Phys. Lett. 116, 204002 (2020).
22. N. Huo, Y. Liu, J. Li, L. Cui, X. Chen, R. Palivela, T. Xie, X. Li, and Z. Y. Ou, Phys. Rev. Lett. 124, 213603 (2020).
23. B. Yurke, S. L. McCall, and J. R. Klauder, Phys. Rev. A 33, 4033 (1986).
24. M. V. Chekhova and Z. Y. Ou, Adv. Opt. Photonics 8, 104 (2016).
25. Z. Y. Ou and Xiaoying Li, APL Photonics, 5, 080902 (2020)
26. D. A. Kalashnikov, A. V. Paterova, S. P. Kulik, and L. A. Krivitsky, Nat. Photonics 10, 98 (2016).
27. A. V. Paterova, and L. A. Krivitsky, "Nonlinear interference in crystal superlattices", Light Sci. Appl. 9, 82 (2020)
28. X. Li, J. Chen, P. Voss, J. E. Sharping, and P. Kumar, Opt. Express 12, 3737 (2004).
29. Yoon-Ho Kim and Warren P.Grice, Opt. Lett. 30, 908 (2005).
30. M. Liscidini, and J. E. Sipe, Phys. Rev. Lett. 111, 193602 (2013).
31. S. D. Dyer, B. Baek, and S. W. Nam, Opt. Express 17, 10290–10297 (2009)
32. J. Li, L. Cui and X. Li, "Two-photon state generated by SU(1,1) nonlinear interferometer: influence of loss," (in preparation).